\documentstyle[12pt]{article}
\newcommand{\lsim}{\mathop{}_{\textstyle \sim}^{\textstyle <}}

\begin{document}
\baselineskip=17pt
\begin{titlepage}
\begin{flushright}
  TU-652\\[-1mm]
  hep-ph/0204293
\end{flushright}

\begin{center}
\vspace*{1.5cm}

{\large\bf 
Dynamical Solution to Supersymmetric CP Problem \\
with Vanishing B Parameter
}
\vspace{8mm}

Masahiro~Yamaguchi and Koichi~Yoshioka
\vspace{2mm}

{\it Department of Physics, Tohoku University, Sendai 980-8578, Japan}\\
\vspace{1.5cm}

\begin{abstract}
The CP violation gives rise to severe restriction of soft breaking
terms in supersymmetric standard models. Among them, constraints on
the holomorphic soft mass of Higgs doublets (the $B$ parameter) are
difficult to satisfy due to the other inherent problem in the Higgs
potential; the $\mu$ problem. In this letter, it is argued that these
CP and $\mu$ problems can be rather relaxed provided that $B$ is
vanishing at high-energy scale. A generic mechanism and some examples
of model are presented to dynamically realize this condition by
introducing gauge singlet fields.
\end{abstract}
\end{center}
\end{titlepage}

\setcounter{footnote}{0}
\noindent
{\sl 1.~Introduction and Idea}
\medskip

Supersymmetry is one of the most attractive extensions of the standard 
model (SM). It has a variety of interesting properties in
phenomenological and theoretical viewpoints, which can bring novel
approaches to unresolved problems of the SM\@. For example, grand
unification with supersymmetry predicts low-energy values of gauge
couplings precisely consistent with the experimental
data~\cite{GU}. In the others, the weak/Planck mass hierarchy is
stabilized against radiative corrections due to the non-renormalizable 
nature of superpotential terms~\cite{NR}.

However, supersymmetry is not a symmetry experimentally observed at
low-energy scale and must be broken by soft breaking terms, which do
not reintroduce quadratic divergences and not spoil mass
hierarchies~\cite{soft}. These soft supersymmetry breaking terms
consist of gaugino masses, scalar trilinear couplings, and scalar
masses. Thus the number of physical couplings is dramatically
increased compared to an exactly supersymmetric case or that of the
SM\@. Moreover, generic forms of soft breaking terms tend to lead
phenomenological disasters such as too large rate of flavor-changing
rare processes like $\mu\to e\gamma$ as well as large CP violation,
which is the subject of this letter. To avoid these problems and make
models viable, supersymmetry-breaking dynamics is required to have
some special properties, and various such mechanisms have been
proposed in the literature. Physical mass spectrum can be explicitly
calculated for a fixed mechanism of supersymmetry breaking and
mediation to the SM sector.

In general, CP-violation measurements provide one of the severest
constraints for supersymmetry-breaking couplings. An important point
is that CP violation occurs even in the absence of flavor
violation. Working with a rather strong hypothesis of flavor
universality for soft terms, there still remain two types of 
CP-violating phases which cannot be rotated away by field
redefinition. One is the phase of scalar trilinear couplings ($A$
terms) relative to that of gaugino masses, and another is that of
holomorphic bilinear terms of scalars ($B$ terms). These phases (in
the basis where gaugino masses are real) are required to be nearly
real by the experimental results such as non-observation of sizable
electric dipole moments of the neutron and leptons~\cite{EDM}. To
satisfy the constraints seems to need some non-trivial, involved
mechanisms or unnatural fine-tuning of model parameters. This is the
CP problem in supersymmetric extensions of the SM~\cite{CP}.

As for $A$ terms, it is found that phase constraints may be somewhat
weak. This is partly because $A$ terms contribute to the electric
dipole moments of leptons, which highly restrict models, only through
the neutralino diagrams with small parameters ($U(1)$ gauge coupling
and a ratio of two Higgs vacuum expectation values
(VEVs)). Furthermore initial complex phases are reduced by 
renormalization-group running down to low energy. From a viewpoint of
model building, however, supersymmetry-breaking $A$ terms are
correlated with Yukawa couplings. It is a non-trivial task to
reproduce the experimentally observed values of fermion mass hierarchy
and CP violation in the Kaon system while there is no new source of CP
violation besides that of Yukawa couplings.

Notice that for high-scale supersymmetry breaking like gravity
mediation, some kind of implements should be usually introduced to
suppress direct couplings between supersymmetry-breaking and visible
sectors. The `separation' of the two sectors is inevitable, for
example, to avoid dangerous flavor-changing operators radiatively
generated in K\"ahler potential. It is found that such suppression
mechanisms also reduce $A$ terms and then their CP-violating
phases. In this letter, we assume this kind of  separation for
simplicity. On the other hand, low-energy supersymmetry breaking
scenarios like gauge mediation~\cite{gauge} also separate the two 
sectors by introducing messenger fields, and predict vanishing $A$
terms at leading order. Accordingly, in both these cases, low-energy
nonzero values of $A$ parameters are generated via renormalization by
gaugino masses. Hence relative phases of $A$ in the basis where
gaugino masses are real turn out to be zero, which is suitable for
obtaining CP-conserving results in supersymmetric SM. 

However the situation is very different for $B$ terms, and naive
separation mechanisms do not work for the CP problem, unlike $A$
terms. In supersymmetric SM, $B$ appears in the holomorphic soft mass
of the Higgs doublets. It is well-known that the most delicate issue
in dealing with the Higgs $B$ parameter is that it closely relates to
other problems in the Higgs potential, which involves the
supersymmetric mass parameter $\mu$\@. The supersymmetric SM Higgs
sector has to satisfy the following three conditions for realizing
proper phenomenology;
\begin{equation}
\mu \;\ll\; \Lambda,
\label{1st}
\end{equation}
\begin{equation}
B\mu \;\lsim\; \mu^2,
\label{2nd}
\end{equation}
\begin{equation}
\arg (M^*B) \;\simeq\; 0,
\label{3rd}
\end{equation}
where $\Lambda$ is a cutoff scale of supersymmetric SM (e.g.\ the GUT
or Planck scale), and $M$ is a universal gaugino mass. We here take
a convention where the $\mu$ parameter is real. The first condition
(\ref{1st}) is nothing but the well-known gauge hierarchy problem
(the $\mu$ problem); it is unnatural to have a tree-level $\mu$
parameter of the electroweak scale which is much smaller than the
natural scale of theory $\Lambda$. A vanishing $\mu$ may be natural by
symmetry argument but has been excluded by the LEP
experiment. Supersymmetry itself can provide a support to stabilize a
tree-level value against quantum correction but does not give any
reasons why $\mu$ is finite and so small relative to larger
fundamental scales. The second condition (\ref{2nd}) is also necessary
for the correct electroweak symmetry breaking. If this condition is
not satisfied, the potential is unbounded from below along some
supersymmetric flat direction. It should be noticed that the condition
(\ref{2nd}) implies a potential difficulty that the
supersymmetry-conserving parameter $\mu$ has to be correlated with the 
supersymmetry-violating one $B$.

The third condition (\ref{3rd}) is concerned with CP violation as
stated above. It says that nonzero phases of a gaugino mass and 
the $B$ parameter must be aligned. One could see whether this
condition is satisfied or not once one knows about all structures of
supersymmetry breaking dynamics. That is, since gaugino masses appear
through a gauge kinetic function and on the other hand, the $B$
parameter resides in the Higgs potential, they apparently have no
connection to each other. Nevertheless Eq.~(\ref{3rd}) imposes a
Fermi-Bose relation even in supersymmetry-breaking sector. Since there
has been few viable mechanisms to arbitrarily control phase values,
phase alignment of the terms which originate from different sources is
rather unnatural and needs some adjustment of model parameters.

In this way, each condition holds in itself independent problems to
solve. Therefore it seems an annoying issue to construct realistic
models satisfying all the requirements on the Higgs potential. In
practice, various attempts have been done to have successful
mechanisms.

For a tree-level $\mu$ term, the required large hierarchy (\ref{1st})
is the foremost problem to be considered. A weak-scale $\mu$ could
arise if some symmetries forbid a bare coupling and small
symmetry-breaking effects generate an effective $\mu$ term. In this
case, generally, a nonzero $B$ parameter is also effectively induced.
However its phase is less under control without any additional
assumptions and may conflict with the constraint (\ref{3rd}) for the
CP problem. As we will see below, $B$ often contains a term
proportional to the gravitino mass $m_{3/2}$. The magnitude of the
gravitino mass is fixed by the requirement that the cosmological
constant should vanish. Its phase is, however, not constrained and is
in general different from that of the gaugino mass. For example, in
the string-inspired supergravity model, the $B$ parameter is given 
by $B=A-m_{3/2}$ in the dilaton dominant case~\cite{sugra}. In this
case, the phase of the $A$ parameter is the same as that of the
gaugino mass and given by that of the dilation $F$ term which is
generally undetermined. Then the phase of $B$ is uncorrelated to the
gaugino mass phase. Other scenarios along this line have also been
discussed in~\cite{others}.

Non-minimal form of Higgs K\"ahler potential is another origin of
the Higgs mass couplings~\cite{GM}. Nonzero $\mu$ and $B$ parameters
are generated via supersymmetry-breaking effects and then cannot be so
hierarchical, which is suitable for the requirement (\ref{2nd}). This
mechanism, however, may seem to need to introduce tuning of parameters
or involved structures of models in order to suppress the CP phase of
the $B$ parameter.

In theories where soft mass terms are generated by loop diagrams
as in gauge mediation or anomaly mediation~\cite{RS,GLMR}, another
severer problem often emerges. This is due to the fact 
that $\mu$ and $B\mu$ are usually generated at the same loop
order. Thus $B=B\mu/\mu$ is given by the supersymmetry breaking scale
without loop suppression which is much larger than other mass
parameters of super-particles including the Higgs doublets. This
conclusion seems quite generic and hard to resolve, and naive attempts
such as a tree-level $\mu$ term or non-minimal K\"ahler potentials
discussed above do not work. Some mechanisms to resolve the difficulty
were discussed in \cite{GM2}.

In this letter, we would like to stress that there is a simple
assumption that rather relaxes these tight constraints on the Higgs
mass parameters. That is, at some high-energy scale, the $B$ parameter
is vanishing;\footnote{This condition was considered in gauge
mediation~\cite{B0other} as well as in gaugino
mediation~\cite{gaugino-m}.}
\begin{equation}
B \;=\; 0.
\label{B0}
\end{equation}
One can immediately see that this assumption resolves the problem
(\ref{3rd}). A low-energy (electroweak) nonzero $B$ term is generated
via renormalization group evolution with the boundary condition
(\ref{B0}). The running of $B$ is driven by gaugino masses and
possibly the large scalar top coupling. The latter is controlled also
by the gaugino masses as long as $A$ terms are supposed to be small at
high-energy scale. This situation is achieved, for example, with the
mechanisms explained before. The gaugino masses thus govern the
low-energy $B$ parameter including its phase, and hence the condition
(\ref{3rd}) turns out to be satisfied. Interestingly, this solution
for the CP problem is independent of actual phase values, which is
difficult to control. Vanishing $B$ is natural in a sense that some
symmetries such as Peccei-Quinn symmetry and possibly $R$ symmetry are
restored in this limit.

The remaining problems of Higgs potential, (\ref{1st}) and
(\ref{2nd}), are simultaneously settled if a nonzero $\mu$ is
generated by supersymmetry-breaking parameters (and can be
real)~\cite{others}. Suppose that there is no $\mu$ term in the vacuum
without supersymmetry breaking, and breaking effects shift the vacuum
resulting in an effective $\mu$ term. The $\mu$ problem (\ref{1st}) is
thus solved in a technically natural way. In addition, $B$ is now
generated by renormalization-group evolution and becomes of the order
of gaugino masses, and the condition (\ref{2nd}) is
achieved. Interestingly, the solution can be discussed only within the
Higgs potential sector and needs not to require knowledge of other
parts of theory such as gaugino masses. It is a reasonable situation
that $B$ is vanishing since mechanisms to control phase values have
not been established.

In this letter, we present a mechanism where the solution with the
boundary condition (\ref{B0}) is dynamically realized. We will discuss
supersymmetry-breaking effects at high-energy scale, but the mechanism 
presented here is general and can be applied to other
supersymmetry-breaking models. Phenomenological implications of this
solution to the supersymmetric CP problem will be discussed
elsewhere~\cite{pheno-CP}.

\bigskip

\noindent
{\sl 2.~Models}
\medskip

As mentioned before, we suppose that a separation of
supersymmetry-breaking and visible sectors occurs. If this is not the
case, induced higher-dimensional K\"ahler terms which link two sectors
generate phenomenologically dangerous operators for flavor-changing
neutral currents, etc. Such an implement is indeed necessary for
models to be viable. A separation may be accomplished, for example, in
a geometrical way in higher-dimensional theories~\cite{RS} or by
strong coupling dynamics of superconformal field
theories~\cite{LS}. The latter might be interpreted as a gravity dual
of the former. In this letter, we assume such mechanisms for
separation, for simplicity, and concentrate on dynamics in the visible
sector. With this locality at hand, in particular, scalar trilinear
supersymmetry-breaking terms vanish at leading order of perturbation
theory. However scalar bilinear terms do not necessarily share the
same result. This is because a non-trivial $\mu$-generation mechanism
has to be fixed as discussed in the introduction.

Let us consider models in which SM gauge singlet fields generate the
Higgs bilinear couplings. Supersymmetry-breaking effect is expressed
in terms of the gravitational multiplet. The relevant part of
Lagrangian is
\begin{eqnarray}
  {\cal L} &=& \int d^4\theta\, \left[\phi^\dagger\phi\,
  f(S,S^\dagger,\cdots) +H^\dagger H+\bar H^\dagger\bar H \right]
  \nonumber \\[1mm]
  &&\qquad\quad +\int d^2\theta\,\phi\, \lambda g(S)H\bar H 
  +\int d^2\theta\, \phi^3\, W(S,\cdots),
  \label{L}
\end{eqnarray}
where $S$ is the singlet which couples to the Higgs doublets through a
function $g(S)$ in the superpotential with a coupling 
constant $\lambda$. Contribution of possible additional (singlet)
fields are denoted by $\cdots$. A chiral superfield $\phi$ is the
compensator multiplet with Weyl weight 1. In the case we consider, a
VEV of $\phi$ is the only source of supersymmetry breaking in the
visible sector and is determined by hidden sector dynamics by
requiring a vanishing cosmological constant,
\begin{equation}
  \phi \;=\; 1+F_\phi \theta^2.
\end{equation}
In the Lagrangian (\ref{L}), we have rescaled the Higgs fields into a
canonical form. It is found convenient to work in this basis, since
the physical masses of Higgs fields are clearly understood and the
effects of $\phi$ to singlet fields is easier to evaluate. With
appropriate form of potentials, nonzero VEVs of singlet fields are
generated after supersymmetry breaking and can provide a solution to
the $\mu$ problem.

The equations of motion for the singlet auxiliary components are
\begin{equation}
  F_\phi\,\partial_{\bar i} f +\sum_j F_j\,\partial_{\bar i}
  \partial_j f +\partial_{\bar i} W^* \;=\; 0,
  \label{F}
\end{equation}
where $\partial_{i}$ ($\partial_{\bar i}$) denotes the derivative
with respect to a field $\Phi_i=S,\cdots$ ($\Phi_i^*=S^*,\cdots$).
Here and in the following, we consider the vacua 
around $H=\bar H=0$. In this case, any details of Higgs fields such as
K\"ahler form are irrelevant to the potential analyses. By integrating
out the auxiliary components and minimizing the scalar potential of
singlet fields, one finds the VEVs of scalar components satisfy the
following equations:
\begin{equation}
  0 \;=\; \frac{\partial V}{\partial \Phi_i} \;=\;
  -2F_\phi\partial_i W +
  \sum_j \partial_i F_j^* \sum_k F_k\partial_k\partial_{\bar j} f,
  \label{V}
\end{equation}
where $F_i$'s have been replaced with the scalar components through
the equations (\ref{F}). The factor $2$ in the right-handed side of
Eq.~(\ref{V}) is related to the fact that superpotential terms have
Weyl weight 3. The Higgs mass parameters $\mu$ and $B$ are given by
\begin{equation}
  \mu \;=\; \lambda g,\qquad -B\mu \;=\; \lambda 
  \bigl( F_\phi g +F_S\partial_S g \bigr).
  \label{muB}
\end{equation}

Here two comments are in order. First it is found from Eq.~(\ref{muB}) 
that a tree-level $\mu$ term, namely, $g=const$ leads 
to $B=-F_\phi$. This is too a large value for electroweak symmetry
breaking to be turned on as long as soft terms are induced at loop
level. Secondly, we do not include a non-minimal Higgs K\"ahler 
term, $\int d^4\theta \frac{\,\phi^\dagger}{\phi}H\bar H$, which could
give another origin of the Higgs mass couplings~\cite{GM}. As
mentioned in the introduction, it does not give a complete solution to
the Higgs mass problems without introducing fine-tuning of couplings
or involved model structures; $\mu\propto F_\phi^*$ and $B=F_\phi$,
and generally, the phases of $B\mu$ and gauginos are different. In the
following, we simply assume the minimal form of K\"ahler 
potential, $\partial_i\partial_{\bar j}f\,=\,\delta_{ij}$ as a good
first-order approximation. Note, however, that even when higher-order
K\"ahler terms suppressed by powers of the fundamental scale are taken
into account, they are irrelevant in the analyses unless models
contain VEVs very close to the fundamental scale.

Our mechanism for a vanishing $B$ term introduces two singlet fields,
here we call $S$ and $N$. Let us consider the following form of
superpotential terms:
\begin{equation}
  \int d^2\theta\,\phi\, \lambda g(S)H\bar H +\int d^2\theta\,\phi^3
  \Bigl[ \lambda' Ng^2(S)+W(S,\cdots) \Bigr],
  \label{superpot}
\end{equation}
where $g(S)$ and $W(S)$ are arbitrary functions of $S$, and $\lambda$
and $\lambda'$ are coupling constants. From the equations of motion
(\ref{F}) and (\ref{V}), one can see that the above superpotential
leads to $B=0$ at the potential minimum with respect to the $N$
field. In fact, Eq.~(\ref{V}) for $\Phi_i=N$ now reads
\begin{equation}
  0 \;=\; \frac{\partial V}{\partial N} \;=\;
  -2 \lambda' g \bigl(F_\phi g +F_S\partial_S g\bigr).
\end{equation}
Then the $B$ parameter (\ref{muB}) vanishes as long as the VEV of $g$
is nonzero, that is, a nonzero $\mu$ parameter is generated. Actual
values of the couplings are irrelevant to the result.

A key ingredient is that the superpotential has at most linear
dependence of $N$. With the above suitable form of the $N^1$ term, the
two equations, the minimization by $N$ (i.e., Eq.~(\ref{V}) with the
minimal K\"ahler) and the $B$ parameter in (\ref{muB}), take the same
form besides nonzero overall factors.\footnote{Even if we worked with a
non-minimal, involved K\"ahler potential $f$, the same result can be
obtained by additional conditions 
for $f$;~ $\partial_N\partial_{\bar S}f=\partial_N^2 f=0$.}
In this case, a separation that suppresses direct couplings between
the hidden and visible sectors plays an important role. For example,
if we have a K\"ahler term
\begin{equation}
  \int d^4\theta\, Z(X,X^\dagger) N^\dagger N,
\end{equation}
where $X$ is the field responsible to supersymmetry breaking in the
hidden sector, it induces a tree-level soft mass of $N$ and may
destabilize our vacuum of $B=0$. Separation mechanisms we assume
throughout this letter is crucial to derive our solution as well as
for suppressing other flavor problems.

Thus the desired boundary condition $B=0$ is dynamically realized
with the superpotential (\ref{superpot}).\footnote{We assume
that there is no tadpole of $N$, whose existence may disturb our
vacuum. Such a tadpole could be prevented by ($R$) symmetries, which
requires some modification of the model. Symmetries would also
restrict the superpotential in desired forms. For example, an $R$
symmetry under which $N$ has a charge $+2$ leads to a linear
dependence of $N$ in the superpotential.}
Notice that $W$ does not play any role in obtaining a vanishing $B$
term. It is clear from the above derivation that the only requirement
which $W$ must satisfy is that it does not depend on the $N$
field. Therefore additional dynamics can be incorporated into $W$ (and
also the K\"ahler of $S$) to have preferable VEVs of singlet fields
while the result $B=0$ is still preserved. Furthermore, as for a
polynomial $g(S)$, it might be curious to have the $Ng^2(S)$ term in
the Lagrangian (\ref{superpot}). However as stressed above, $B=0$ is
guaranteed as long as the $W$ part does not contain the $N$
field. Accordingly it is rather easy to have $Ng^2(S)$ with polynomial
$g$ if we start with the Lagrangian
\begin{equation}
  \int d^2\theta\,\phi\,\lambda TH\bar H + \int d^2\theta\,\phi^3
  \Bigl[ \lambda' NT^2 + \kappa U (T-g(S)) + W(S,\cdots) \Bigr],
\end{equation}
where $T$ and $U$ are the gauge singlets. With this action, $B=0$ is
also achieved. On the other hand, integrating out $T$ and $U$, we have
just the Lagrangian (\ref{superpot}) with polynomial $g$. It could be
easier for the above form of action to follow from symmetries.

With the condition $B=0$ at hand, a low-energy $B$ term is generated
by gaugino mass effects in renormalization group evolution. The phase
of $B$ is therefore automatically aligned with that of gaugino masses, 
and suppression of CP violation is accomplished. Another part of our
solution to the Higgs mass problems is that the $\mu$ term is induced
by supersymmetry breaking dynamics. In the present case, this
corresponds to the effects of the compensator field $F_\phi$ in the
visible sector. The VEV of the scalar component of $S$, which gives
rise to the $\mu$ parameter, is fixed by minimizing potential with
respect to $S$. The $\mu$ problem (\ref{1st}) can be solved with an
appropriate form of the functions $g(S)$ and $W(S)$. Deviations from
the minimal K\"ahler form of $S$ could also stabilize the $S$
field. It should be noted that detailed forms of $g$ and $W$ do not
affect the above argument for the vanishing $B$ parameter.

\bigskip

Let us study several examples of $g$ and $W$, and estimate orders of
magnitude of effectively induced $\mu$ parameter. Our aim here is not
to present complete theories but to give simple toy models in order to
study structures of the mechanism. At first, as a special case,
consider $g=S$ and $W=S^3$, that is, all the superpotential terms are
cubic and renormalizable ones. In this case, however, the VEV of $S$
is undetermined by the supersymmetry breaking $F_\phi$. This is
because the compensator $\phi$ can be absorbed by the 
rescaling $S\phi\to S$, $\cdots$, and hence does not give
supersymmetry-breaking potential at tree level. Therefore dimensionful
parameters and/or higher-dimensional operators have to be introduced
in the Lagrangian.

\medskip

\noindent
[ Model 1 ]
\smallskip

\noindent
First we consider the model
\begin{equation}
  g \;=\; S,\qquad W \;=\; m_S\, S^2,
\end{equation}
where $y$ is the coupling constant of $O(1)$. A tree-level $\mu$ term
may be forbidden by imposing a discrete 
symmetry $S\to -S$ and $\bar H\to -\bar H$. This symmetry also forbids
a generation of tadpole operator for the $S$ field as well as the
dimension 5 operators for nucleon decay. In the supersymmetric 
limit $F_\phi=0$, $S$ is forced to be zero. Then at the minimum 
of the potential including supersymmetry breaking, the VEV 
of $S$ flows to a nonzero 
value $\langle S\rangle\simeq\frac{1}{\lambda'} (m_S F_\phi)^{1/2}$. 
The VEV of $N$ is determined by the equation of motion
for $N$ and is given 
by $\langle N\rangle\simeq\frac{1}{\lambda'}m_S$. We note that
around this vacuum, radiatively induced soft masses are negligibly
small and do not disturb the potential analysis here. Integrating out
the high-energy dynamics, the effective $\mu$ parameter is generated
as
\begin{equation}
  \mu \;\simeq\; \frac{\lambda}{\lambda'}(m_S F_\phi)^{1/2},
  \label{mu1}
\end{equation}
which is a geometric mean of $m_S$ and the gravitino mass $F_\phi$.

The scale of $\mu$ depends on both $m_S$ and $F_\phi$. In the case of
high-energy supersymmetry breaking, i.e.\ the heavy gravitino, $m_S$
must be chosen as a TeV scale for a small $\mu$
parameter.\footnote{Alternatively one can take $\lambda$ small enough
to make $\mu$ in the correct order of magnitude.  This does not,
however, provide a natural explanation of the $\mu$ parameter.} 
A natural way to achieve this is to slightly modify the K\"ahler form
of $S$~\cite{GM}. If K\"ahler potential contains the quadratic term 
of $S$, $\int d^4\theta\phi^\dagger\phi S^2+{\rm h.c.}$, it induces a
mass $m_S\simeq F_\phi^*$. (This deformation of K\"ahler potential
also induces a holomorphic soft mass of $S$ but does not affect the
result (\ref{mu1}).) \ Thus $\mu$ is roughly on the correct order of
magnitude; $\mu\simeq F_\phi$. When soft terms are loop induced as in
anomaly mediation, some tuning of couplings is still required. Other
mechanisms discussed in the literature~\cite{others} can also be
utilized to obtain a supersymmetry-breaking order of mass $m_S$.

For low-energy supersymmetry breaking, a gravitino mass is much
smaller than the weak scale, $F_\phi\sim F_{\rm hid}/M_{\rm Pl}$. 
Here $\sqrt{F_{\rm hid}}$ is a supersymmetry-breaking scale in hidden
sector which is separated by taking $\sqrt{F_{\rm hid}}$ as much lower
than the fundamental scale. As a consequence, $A$ terms are
reduced. From Eq.~(\ref{mu1}), in this case, a correct order of $\mu$
parameter is realized if $m_S$ is slightly lower than the Planck
scale. For example, with $\sqrt{F_{\rm hid}}\sim 100$ TeV, $m_S$ is
around the GUT scale $\sim 10^{16}$ GeV for a weak-scale $\mu$
parameter. This example gives a solution to the $\mu$ problem in gauge
mediation scenarios. The well-known trouble of too a large $B$
parameter is avoided in the present case by realizing the 
condition $B=0$ dynamically. The $\mu$ term is generated at tree level
but the $B$ term at loop level.

\medskip

\noindent
[ Model 2 ]
\smallskip

\noindent
Next let us discuss an example with higher-dimensional operators
(superpotential terms whose mass dimensions are greater than three).
The superpotential is given by
\begin{equation}
  g \;=\; S^n,\qquad W \;=\; yS^{n+2},
\end{equation}
where the exponent $n$ is not necessarily an integer but 
satisfies $n>1$ unless the coupling $\lambda'$ is too 
small. (The $n=1$ case corresponds to the conformal limit discussed
above where $S$ is not determined by supersymmetry-breaking effects.)
The $S$ direction is stabilized by the potential also in this
case. Minimizing the potential, the $\mu$ parameter in the vacuum is
found to be
\begin{equation}
  \mu \;\simeq\; \frac{1-n}{2n(n+2)}\frac{\lambda}{y}\,F_\phi.
\end{equation}
We find that $\mu$ is proportional to $F_\phi^{\,1}$ irrespective 
of $n$, and therefore this model gives a solution to the $\mu$ problem
relevant to high-energy supersymmetry breaking like gravity mediation
and other scenarios where the gravitino mass is around the electroweak
scale~\cite{other}. This example shares essential features with the
Model 1. For example, a tree-level $\mu$ can be forbidden with a
discrete symmetry, $S^n\to -S^n$ and $\bar H\to -\bar H$ (that is
broken by higher-dimensional operators).

In case of $n\gg 2$, the singlet $N$ tends to have a rather flat
potential and a large value of VEV beyond the Planck scale, though
depending on the couplings. A natural way to cure this problem is to
add a small soft mass for $N$, which stabilizes the 
VEV $\langle N\rangle$ to an intermediate scale value irrespectively
of $n$. The small soft mass, roughly around a MeV to GeV scale, can be
induced by a small deviation from the minimal K\"ahler form (the exact
separation) without conflicting with flavor-changing neutral current
constraints. It is also found that introducing soft mass of $N$ makes
the model cosmologically safe, with small decay rate and a negligible
contribution to energy density by coherent oscillation of the
scalar. We note that the problem could also be removed by adopting
other forms of potential $W$ including relevant fields.

\bigskip

\noindent
{\sl 3.~Summary}
\medskip

We have shown that the condition $B=0$ for a solution to the
supersymmetric CP problem can be dynamically realized in the Higgs
potential with gauge singlet fields. With the appropriate terms in
superpotential, minimizing scalar potential with respect to singlet
fields exactly leads to a vanishing $B$ parameter. At the same time,
the supersymmetric mass $\mu$ is generated by supersymmetry-breaking
effects. A nonzero $B$ parameter is induced during
renormalization-group evolution down to low energy. It is driven by
gaugino masses and hence the CP-violating phase of $B$ parameter
vanishes in the basis where gaugino masses are real. This solution to
the CP problem does not need any information about gaugino masses and
can be discussed only in the Higgs sector. Our solution can be
incorporated in various mediation mechanisms, including gauge
mediation, gaugino mediation, etc. Vanishing $A$ and $B$ terms is
shown to be an attractive way to solve the CP and $\mu$ problems and
be attainable with natural physical implications.

\vspace*{5mm}
\subsection*{Acknowledgments}

We would like to thank N.~Maekawa and T.~Takahashi for helpful
discussions and comments. This work was supported in part by the
Grant-in-aid from the Ministry of Education, Culture, Sports, Science
and Technology, Japan (No. 12047201) in part by the Japan Society for
the Promotion of Science under the Postdoctoral Research Program
(No.~07864).


\end{document}